\documentclass[12pt]{article}
\usepackage{graphics,epsfig}
\def\abs#1{\left\vert #1 \right \vert}

\def\Dron#1#2{{\Frac{\partial#1}{\partial#2}}}

\def\Dd#1#2{{\Frac{\d#1}{\d#2}}}

\def\d{{\hbox{\rm d}}}

\def\frac#1#2{{\textstyle{{#1} \overwithdelims.. {#2}}}}
\def\Frac#1#2{{\displaystyle{{#1} \overwithdelims.. {#2}}}}
\def\$#1${{
\begin{flushleft}
\lineskip=3D3pt \lineskiplimit=2pt
\tolerance=7000
$\displaystyle #1$
\lineskip 1.5pt \lineskiplimit=1pt
\end{flushleft}
}}

\evensidemargin=\oddsidemargin

\def\[#1\]{\begin{eqnarray} #1 \end{eqnarray}}
\begin{document}

\title{Solving the Triangular
Ising Antiferromagnet \\by \\Simple Mean Field}
\author{
Serge Galam$^1$ and Pierre-Vincent Koseleff$^2$\\\\
$^1$Laboratoire des Milieux D\'{e}sordonn\'es et
H\'et\'erog\`enes
\thanks{
Laboratoire associ\'e au CNRS (UMR n$^{\circ}$ 7603), galam@ccr.jussieu.fr
},
Case 86,\\
Universit\'e Pierre et Marie Curie,
4, place Jussieu, F-75252 Paris Cedex 05.
\\\\
$^2$\'Equipe ``Analyse Alg\'ebrique'', Institut de Math\'ematiques
\thanks{
Laboratoire associ\'e au CNRS (UMR n$^{\circ}$ 7586), 
koseleff@math.jussieu.fr},
Case 82\\
Universit\'e Pierre et Marie Curie,
4, place Jussieu, F-75252 Paris Cedex 05.
}
\date{February 20, 2001}
\maketitle

\begin{abstract}

Few years ago, application of the mean field Bethe scheme on a given 
system was shown
to produce a systematic change of the system intrinsic symmetry. For 
instance, once
applied on a ferromagnet, individual spins are no more equivalent. 
Accordingly a new
loopwise mean field theory was designed to both go beyond the one site Weiss
approach and yet preserve the initial Hamitonian symmetry. This 
loopwise scheme is
applied here to solve the Triangular Antiferromagnetic Ising model. 
It is found to
yield Wannier's exact result of no ordering at non-zero temperature. 
No adjustable
parameter is used. Simultaneously a non-zero critical temperature is 
obtained for
the  Triangular Ising Ferromagnet.  This simple mean field scheme 
opens a new way to
tackle random systems.

\end{abstract}

\newpage
\section{Introduction}

Collective phenomenon are rather difficult to solve exactly. Up to date,
only some one dimensional problems and the square zero field
Ising model allow an exact analytical solution [1].
To compensate this situation,
a rich family of approximate methods has been developed
over the last one hundred years. The most
powerful one being the renormalization group techniques [2].

At start was the
Mean Field Theory (MFT). It offers a very practical and simple
tool to solve most collective phenomena [1]. While it is completely universal
and generic, associated quantitative results are unusually poor.
In particular critical temperatures and
exponents are rather far from exact estimates [2].
Sometimes even the order of the transition may be wrong like for the
instance in the Potts model [3].

The crudest and most simple version of MFT is the 1907 Weiss pioneer model
[4].  It reduces the infinite number of fluctuating degrees of freedom down
to one, $S_i$, which couples to homogeneous mean field degrees of freedom
$m$. The thermodynamics is then solved calculating the associated partition
function from which the self-consistent equation
$<S_i>=m$ (where
$<...>$ means thermal average) is derived.

In the case of Ising systems with $q$ nearest neighbor interactions,
Weiss theory gives $<S_i>=tanh (K q m)$
where $K\equiv \beta J$, $J$ is the exchange coupling, $\beta\equiv
\frac{1}{k_BT}$, $k_B$ is the Boltzman constant and
$T$ is the temperature.
Associated critical temperature is $K_c=\frac{1}{q}$.
At odd with the known exact result a phase
transition is obtained at $d=1$ ($q=2$) [1].

 From there it took 28 years before Bethe improved the Weiss model [5].
Instead of just one fluctuating spin, he considers a cluster of fluctuating
spins with a central one and its nearest neighbors.
The main achievement of the Bethe approximation is to yield the exact
result of no ordering at one dimension. However, critical temperatures
given by $K_{c}= \tanh^{-1} (\frac{1}{q-1})$, are not much better than
from Weiss
model. Critical exponents stay unchanged. Latter on, using computer
capabilities,  larger size fluctuating clusters have been considered to obtain
better critical temperatures [6].

However, few years ago the Bethe cluster scheme was showed to
systematically  change the system intrinsic symmetry [7]. Starting 
from a system
with equivalent sites like for instance a square Ising Ferromagnet, 
It ends up making
individual sites unequivalent. At this stage it is worth to stress that an
approximation can be very crude and yet not wrong as long as it preserves
the intrinsic symmetry of the problem. Otherwise it does change the 
physics of it.
We are not talking here about a symmetry breaking of the higher phase 
symmetry as it
occurs in a usual phase transition but of a change of the symmetry of 
the disorder
phase itself.

On this basis the challenge was to find out if it is indeed possible 
to build a MFT
which considers more than two fluctuating spins, yet preserving the 
initial lattice
symmetry. Indeed, Galam showed it is possible using a loopwise scheme 
(LWS) which
articulates on finite-size one-dimensional closed loops [7].
Paving the
whole lattice with these loops, half of them are kept fluctuating 
while the other
half is averaged out with mean field degrees of freedom. The scheme 
is illustrated in
Figure 1 for the square lattice.

\begin{figure}
\begin{center}
\caption{ The loopwise scheme in the square case: s1, s2, s3, s4 are the
fluctuating spins while m1, m2, m3, m4 are mean field averages}
\epsfbox{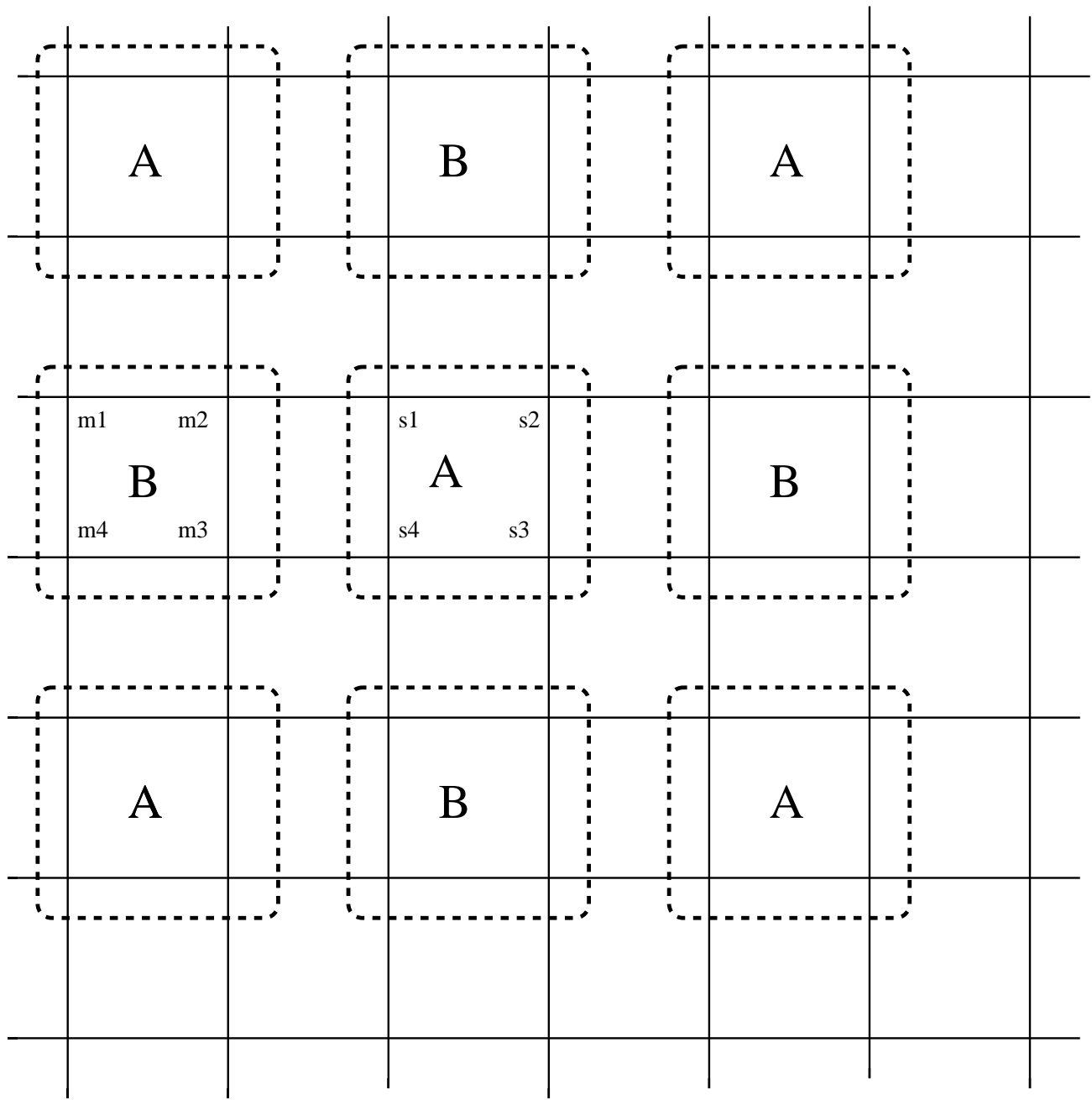}
\end{center}
\end{figure}

The LWS is a generic model. It was applied to a large class of
ferromagnetic systems on Bravais lattices [7, 8]. It reproduces the
exact result of no ordering at one dimension. Moreover, for Ising 
hypercubes, it
exhibits a lower critical dimension $d_l$ for long range ordering 
which is equal
to the Golden number $d_l=\frac{1+\sqrt 5}{2}$. However critical exponents are
unchanged from Weiss model.

On this basic, to determine the range of validity of this new LWS, it 
is of interest
to check if it can yield new properties which are out of reach of 
previous mean field
theories like frustration.  For instance, when applied to the fully-frustrated
Triangular Ising Antiferromagnet (TIA) most MFT predict a transition 
at a non-zero
temperature while  an earlier exact argument by Wannier proved no 
symmetry breaking
occurs at any non-zero temperature [9].

Few years ago, to bridge this difficulty Netz and Berker introduced
the hard spin recipe [10]. It combines
a mean field calculation with some Monte Carlo sampling. When applied
to the TIA,
it yields the correct result of no
ordering at $T\neq 0$.
Later Banavar et al suggested that the Monte Carlo
sampling could be reproduced by
expanding  all possible products of the 6 nearest neighbors spins of the
``exact spin'' but it was then disproved by Netz and Berker[11].

More recently focusing on the TIA, Monroe approximated
the triangular lattice
with a Husimi tree built up of triangles [12]. It then allows to
include properly frustration to get a
correct phase diagram. However an Huzimi tree
is not a triangular lattice.

In this paper we apply the very simple LWS to the
fully frustrated Triangular Ising Antiferromagnet (TIA).
The Wannier exact result is recovered [10] and a transition is found at $T=0$.
The following of the paper is organized as follows. Section 2 deals with
the frustration effect. In Section 3 the LWS is presented. The TIA is solved
analytically in Section 4 using the LWS. In Section 5 using the same Equations,
the Triangular Ising Ferromagnet (TIF) is also solved. Some possible 
applications are
mentinned in the Last Section.

\section{The Frustration effect}

Frustration is a major ingredient of many physical systems. It results
from the impossibility to minimize simultaneously all pair interactions.
In turn it makes the ground state highly degenerate [9]. Frustration
effects may arise from either quenched disorder or topological
constraints.

Random bond spin glasses
are the archetype of frustration produced by disorder. The random
distribution
of quenched competing interactions generates analytical difficulties
in calculating the thermodynamic functions. In particular to  average
the disorder over the logarithm of the partition function is yet a real
theoretical challenge. Usual mean field treatments failed to incorporate
simultaneously frustration and quenched randomness.

On this basis the TIA has the advantage of being fully frustrated without any
disorder making the study of frustration itself more easy. It is therefore the
perfect candidate to check the ability
of a new scheme to deal with frustration.
In addition an earlier exact argument by Wannier [10] has proved the
absence of symmetry breaking at any non-zero temperature for this system.
At constrast most mean field like approaches produce wrongly some 
non-zero critical
temperature. Along this line, Netz and Berker recipe [10] with Banavar et al
reformulation [11] stand at odd.

\section{The loopwise scheme (LWS)}

The LWS was introduced few years ago to overpass the symmetry inconsistency
of the Bethe scheme, yet retaining its physical feature of including several
fluctuating degrees of freedom [7].

To implement the LWS on any lattice requires to single out two identical
interpenetrating sublattices. Each element being composed from a closed
compact loop of degrees of freedom. The shape and number of these degrees of
freedom are determined by the lattice topology. It is the smallest closed
linear loop. For instance in the square case (Fig. 1) it includes 4 spins
while for the triangular lattice (Fig. 2) 3 spins are involved.
One of the sublattice is fluctuating and the other one is mean field.

Both sublattices are coupled via nearest neighbor
interactions. The problem is thus mapped onto decoupled
one-dimensional closed fluctuating chains in external fields.
The
fields originate from the coupling to the mean field loops.
At this stage an exact analytical calculation can be performed
whatever the chain size is. It is worth to note no adjustable parameter is
used.

The LWS is a generic model. It was applied to a large class of
ferromagnetic systems [7, 8].
Being built on using closed linear loops it should be well adapted to embody
frustration effects [9].

\section{Solving the Triangular Ising Antiferromagnet}

We now apply the LWS to the fully frustrated TIA.
We first partition
the triangular lattice into two interpenetrated triangular sublattices
$A$ and $B$. Thermal
fluctuations are then ignored on the B-sublattice
while preserved within the A-sublattice. These triangles
are closed loops with no center (see Figure 2).

\begin{figure}
\begin{center}
\caption{ The loopwise scheme in the triangular case: s1, s2, s3 are the
fluctuating spins while 1, 2, 3 represent mean field averages m1, m2, m3}
\epsfbox{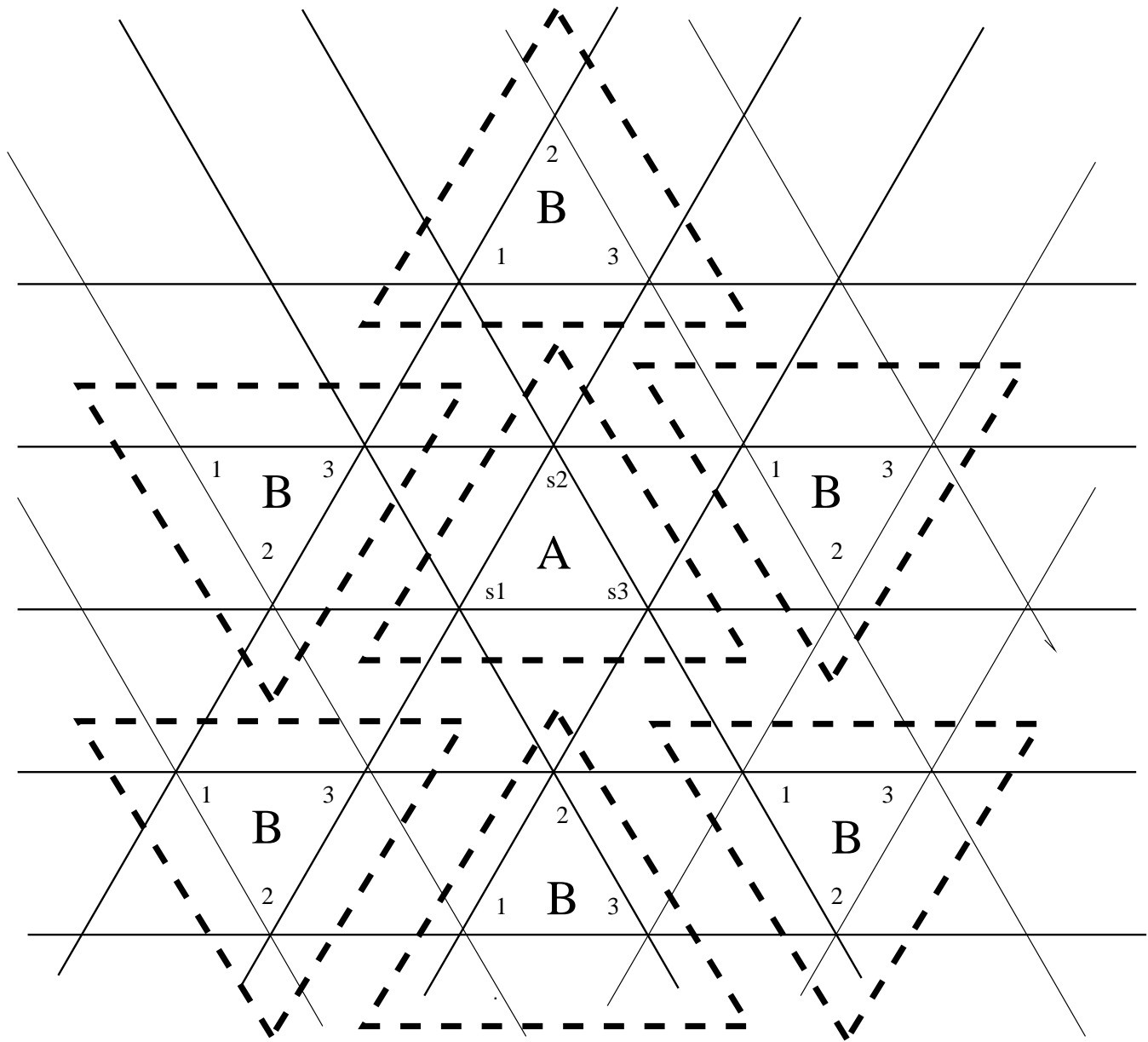}
\end{center}
\end{figure}

All nearest neighbor  (nn)
plaquettes of a A-plaquette are B plaquettes, and vice versa.
Therefore, on a given plaquette each spin has
two nn spins of the same
species (within the same plaquette), and four nn spins of the
other
species (belong respectively to three different nn plaquettes).

Above breaking of the initial lattice symmetry makes
the partition function calculable by decoupling the fluctuating
triangles. The A-sublattice degrees of
freedom can thus be integrated out in the partition function. The initial
lattice symmetry will be restored latter using the usual mean field
self-consistent constraint (Eq. (\ref{eq1}) below).

\subsection{Setting the Equations}

Given a A plaquette, we
label the
3 fluctuating spins $S_1,\,
S_2,\, S_3$. We then
introduce 3 magnetizations $m_1,\, m_2,\, m_3$ for corresponding
B plaquettes (Fig. 3).
The Hamiltonian then writes,
\begin{equation}\label{ham}
H=-J(S_1S_2+S_2S_3+S_3S_1)-\delta J\left (
S_1(m_2+m_3)+ S_2(m_3+m_1)+S_3(m_1+m_2)\right ),
\end{equation}
where $\delta =2$ accounts for the coupling to the B mean field plaquettes.
 From Eq. (\ref{ham}) the partition function is
\begin{equation}\label{Z}
Z=\sum_{S_i=\pm 1}\exp\{-\beta H\} ,
\end{equation}
where $i=1,\ 2,\ 3$.
The three thermal average of $S_1,\, S_2,\, S_3$ are given by,
\begin{equation}
<S_i>=\Frac 1Z \sum_{S_j=\pm 1} S_i \exp\{-\beta H\}\ .
\end{equation}
We can thus write the associated three self-consistent equations,
\begin{equation}
<S_i>= m_i\ .
\label{eq1}
\end{equation}

\subsection{Looking for minima}

Indeed we are looking for minina of the free-energy which results from
the
partition function $Z$. It is then worth to stress not all solutions of
Eq. (\ref{eq1}) are minimum.
A criterium to make Eq. (\ref{eq1}) a derivative of a
function is to require its cross derivatives
with respect to the $m_i$ to be equal, i. e.,
\begin{equation}
\Dron{}{m_{j}} <S_{i}> = \Dron{}{m_{i}} <S_{j}>,
\label{cross}
\end{equation}
for $i,\ j=1,\ 2,\ 3$
\def\Z{\hbox{\bf Z}}
\def\m{\hbox{\bf m}}
\def\S{\hbox{\bf S}}

Writing,
$
\S = (<S_{1}>, <S_{2}>, <S_{3}>), and \ \m = (m_{1}, m_{2}, m_{3}),
$
the problem is now to find a set
$
S = \{ \m \in {\bf R}^{3}; \, \S(\m) = \m \},
$
such that there exists a function $F$ obeying to,
\begin{equation}
(\m -\S(\m))= \hbox{d} F (\m) = 0 \ .
\end{equation}
To solve it, we rewrite thermal averages $<S_i>$ as,
\begin{equation}
<S_{i}> = \Frac 1Z \sum_{s_{i}=\pm 1} s_{i} f(s_{1},s_{2},s_{3})\ ,
\end{equation}
where
\begin{equation}
f (s_{1},s_{2},s_{3})=\exp\left
\{
K (s_{1}s_{2}+s_{1}s_{3}+s_{2}s_{3}) + \delta  K
(m_{1}(s_{2}+s_{3})+m_{2}(s_{1}+s_{3})+m_{3}(s_{1}+s_{2}))\right \}\ .
\end{equation}
Let $\sigma \in \Sigma_{3}$ be a permutation. Considering
$
\sigma(\m)=(m_{\sigma(1)},m_{\sigma(2)},m_{\sigma(3)})
$
we have,
\begin{equation}
\sigma(\Z(\m)) = \Z(\sigma(\m)), \Z(-\m) = -\Z(\m)\ .
\end{equation}
Writing $X = \exp{K}$ and $x_{i}= \exp{\delta K m_{i}}$,
$<S_{i}>$ are rational fractions in $(x_{i},\ X)$ and we have,
\[
Z &=& \Frac{D}{XT_{3}^{2}},
\]
\[
D &=& (1+T_{3}^{2})X^{4} + T_{2}+T_{1}T_{3},
\]
\[
<S_{i}>& =& 1 - 2 \Frac{x_{i}(T_{1}+T_{3}-x_{i})+X^{4}}{D},
\]
where
\[
\begin{array}{rcl}
T_{1} = x_{1}+x_{2}+x_{3}, \\ \\
T_{2}=x_{1}x_{2}+x_{1}x_{3}+x_{2}x_{3}, \\ \\
T_{3} = x_{1}x_{2}x_{3},
\end{array}
\]
are the elementary symetric functions. Note $D>0$, $X>0$, $x_{i}>0$
and $\abs{<S_{i}>} < 1$.

Solving first the $K=0$ case, we get immediately $<S_i>=0$ and the
solution is $m_i = 0$. We can then proceed assuming $K\not =0$.

\subsection{The  most general solution $m_1\not =m_2\not =m_3$}

We can now solve the equations, starting with the most general case $m_1\not
=m_2\not =m_3$.
Eq. (\ref{cross}) is equivalent to,
\[
\Dron{<S_{i}>}{x_{j}}\Dron{x_{j}}{m_{j}} =
\Dron{<S_{j}>}{x_{i}}\Dron{x_{i}}{m_{i}},
\]
that is,
\[
\begin{array}{rcl}
\left (x_1-x_2\right )\left (X^4x_3^2x_2^3x_1^3-x_1^2x_2^2+2 X^
4x_1^2x_2^2x_3^2+x_3^2x_2
x_1-2 X^4x_1 x_2 -X^4\right ) &=& 0, \\ \\
\left (x_1-x_3\right )\left (X^4x_2
^2x_3^3x_1^3-x_3^2x_1^2+2 X^
4x_1^2x_2^2x_3^2+x_2^2x_3
x_1-2 X^4x_3 x_1 -X^4\right ) &=& 0,\\ \\
\left (x_2-x_3\right )\left
(X^4x_1^2x_2^3x_3^3-x_2^2x_3^2
+2 X^4x_1^2x_2^2x_3^2
+x_1^2x_3 x_2
-2 X^4x_3 x_2
-X^4
\right ) &=& 0.
\end{array}
\label{EEE}
\]

Suppose first, three different values of $m_i$. It makes $x_1\neq x_2 \neq x_3$
since $K\neq 0$,
which in turn, solving Eq. (\ref {EEE}) implies,
\[
T_{1} = -\Frac{T_{2}}{T_{3}} \Frac{2T_{3}^{2}-1}{T_{3}^{2}-2}\  ,
\]
and,
\[
X^{4} = \Frac{T_{2}}{T_{3}^{2}-2}.
\]
In conclusion
\[
D  =
(1+T_{3}^{2})X^{4} + T_{2}+T_{1}T_{3},
= 0
\]
which is impossible since $D>0$. Therefore, we can conclude that out of
the
three
$m_i$, two must be equal.
We then suppose $m_{1}=m_{2}\not = m_{3}$.

\subsection{The solution exhibits the symmetry $m_1 =m_2\not =m_3$}

 From  above calculation we restrict the minima search to the
subspace of solution $m_1 =m_2\not =m_3$.
It implies $x_{1}=x_{2}$ and,
\[
<S_{1}>-<S_{3}> &=& -2 \Frac{x_{2}(1+x_{1}x_{3})}{D}
(x_{1}-x_{3}),\label{Kn}
\]
so it makes,
\[
\Frac{\exp(\delta K m_{1}) - \exp(\delta K m_{3})}{m_{1}-m_{3}}<0,
\]
which in turn makes $K<0$.
Let us define $P\equiv x_{2}x_{3}$ and $N\equiv x_{2}^{2}x_{3}$,
it gives,
\[
X^4 =
\Frac{P^{3}-N^{2}}{P(N^{2}P+2N^{2}-2P-1)} \label{x4},
\]
and,
\[
<S_{1}>+<S_{2}>+<S_{3}> = \Frac{P -1}{P+1}. \label{sn}
\]
If $P=1$ or $N=1$ then $m_1=m_2=m_3=0$ which is not possible
since we assumed above $m_{1}=m_{2}\not = m_{3}$. So $(P-1)(N-1) \not  =
0$.
On the other hand, as $K<0$, we must have,
\[
\Frac{1 - N}{m_1 + m_2 + m_3} >0,
\]
which makes $(P-1)(N-1)<0$ but Eq. (\ref{x4})  gives,
\[
X^4 = \Frac 1P
\Frac{P^{3}-N^{2}}{(1-P) + (2+P)(N^{2}-1)} <0,
\]
which is impossible.
In conclusion all three $m_i$ must be equal.
On this basis we now assume $m_{1}=m_{2}=m_{3}=m$.

\subsection{The solution is fully symmetrical with $m_1 =m_2 =m_3=m$}

We have now proved the minima belong to the solution subspace defined by the
symmetry condition $m_1 =m_2 =m_3=m$.
On this basis, writing
$Y = \exp(\delta K m)$, Eq. (\ref{eq1}) becomes,
\[
m = f_{K}(m)  =
{\Frac {\left (Y^{4}-1\right )
\left (\left ({Y}^{8}+{Y}^{4}+1\right ){X}^{4}+{Y}^{4}\right )}{\left
({Y}^{4}+1\right )\left (\left ({Y}^{8}-{Y}^{4}+1\right ){X}^{4}+3\,{Y
}^{4}\right )}},
\]
so we have either $m=0$, or both,
\[
\Frac{Y^{4}-1}{m} = \Frac{e^{4 \delta K m}-1}{m}>0,
\]
and $K>0$. We also deduce  that $\abs{f_{K}(m)} < 1.$

As $f_{K}(-m) = - f_{K}(m)$ it is enough to solve the case $m \geq 0$.
We thus obtain $X>1, Y>1, K>0$, or $m=0$. Computing the derivative gives,

\[
f_{K }'(m) & = &
8\delta K \,\Frac {{Y}^{4}\left (3\,{Y}^{8}{X}^{8}+\left (
{Y}^{16}+4\,{Y}^{12
}+4\,{Y}^{4}+1\right ){X}^{4}+3\,{Y}^{8}\right )}
{\left ({Y}^{4}+1
\right )^{2}\left (\left ({Y}^{8}-{Y}^{4}+1\right ){X}^{4}+3\,{Y}^{4}
\right )^{2}},
\]
\[
f_{K }''(m) &=& -32 \delta^2 K ^{2} \Frac{Y^{4}(Y^{4}-1)}
{\left ({Y}^{4}+1
\right )^{3}\left (\left ({Y}^{8}-{Y}^{4}+1\right ){X}^{4}+3\,{Y}^{4}
\right )^{3}}
\,
g_{K }(m),
\]
where
\[
g_{K }(m) &=&
({Y}^{4}+1)^{6} +(2\,{Y}^{8}+7\,{Y}^{4}+2)
({Y}^{4}+1)^{4}({X}^{4}-1) \nonumber \\
& & +({Y}^{16}+7
\,{Y}^{12}+21\,{Y}^{8}+7\,{Y}^{4}+1)({Y}^{4}+1)^{2
}({X}^{4}-1)^{2} \nonumber \\
&&+9\,{Y}^{8}({Y}^{8}+{Y}^{4}+1
)({X}^{4}-1)^{3} >0.
\]

Therefore when $m \geq 0$, in addition to  $0\leq f_{K }(m)<1$,
we have $f_{K}'(m)>0$ and $f_{K}''(m)<0$. These properties allow to conclude
that,
\begin{enumerate}
\item If $f_{K }'(0)\leq 1$,   $0$ is the only fixed point of
$f_{K }$.
\item If $f_{K }'(0)>1$, $f_{K }$ has exactly three fixed points,
$0, b, - b$ where $-1<b<1$
\end{enumerate}
Computing then,
\[
f_{K }'(0) = 2\delta K \,\Frac{3\exp(4K) + 1}{\exp(4K)+3},
\]
it appears to be an increasing function of $K$. It makes,
\[
f_{K}'(0)=1,
\label{K}
\]
to have a unique solution $K_{0}$.
Moreover, if $K<K_0$ then $ f_{K }'(0) <1$ and
if $K>K_0$ then $ f_{K }'(0) >1$.

\subsection{The actual minima}

Looking for minima of $F_{K }$ where,
\[
\Dd{F_{K }(m)}{m} =
m- f_{K }(m)\ ,
\]
depending on the value of $K$, two cases appear quite naturally for $K >K_{0}$
and $K \leq K_{0}$. It shows the Triangular Ising both Anti and
Ferromagnets are solved simultaneously.

\subsubsection{First case: $K >K_{0}$}

In this case, $f_{K }(m) = m$ has 2 solutions   $m=0$ and
$m^2=a$ where $a$ is a positive function of $K$.
Having,
\[
F_{K }''(m=0) = 1-f_{K }'(m=0)<0\ ,
\]
$m=0$ is a maximum for $F_{K}$.
In parallel,
\[
F_{K}''(m=\sqrt {a}) = F_{K}''(m=-\sqrt {a}) = 1 -
f_{K}'(m_{1}) > 0\ .
\]
Therefore  $m=\sqrt {a}$ and $m=-\sqrt {a}$ are minima of $F_{K}$.
They
correspond
to the Triangular Ising Ferromagnet symmetry breaking at low temperatures where
$K_{0}$ is the associated critical temperature.

\subsubsection{Second case: $K \leq K_{0}$}

Then the unique solution of $f_{K }(m)=m$ is $m=0$.
There,
\[
F_{K }''(0) = 1- f_{K }'(0)>0,
\]
so it is a minimum for $F_{K}$.
This case embodies indeed two different physical situations.
\begin{enumerate}

\item The first range of positive $K$,
$0\leq K \leq K_{0}$,  corresponds to the disordered phase of above Triangular
Ising Ferromagnet.

\item At the same time, the range of negative $K$ ($K\leq 0$) corresponds
to the Triangular
Ising Antiferromagnet. For this system the unique solution is always $m=0$ for
the whole range of temperatures $T>0$. We conclude  situation  proving thus no
It means no ordering occurs for the TIA at any non zero temperature. 
The Wannier
argument is  thus recovered [10].

\end{enumerate}

\subsection{A transition at $T=0$}

 From the exact Wannier solution the Triangular
Ising Antiferromagnet is known to exhibit a phase transition at $T=0$ 
to an ordered
phase with broken symmetry among the three sublattices. Accordingly 
we now examine
what our scheme yield in the case $K \rightarrow\infty$. To solve the 
equations it
is more convenient to rewrite
$Z$ and $<S_{i}>$ in terms of
$T=\tanh(K)$ and
$t_{i}=\tanh(\delta K m_{i})$. We first note $\abs{<S_{i}>} \leq 1$ since
$\abs{\sinh(x)} \leq
\cosh(x)$.
Then, once the $m_{i}$ are fixed within $[-1,1]$, the condition
$K\rightarrow \infty$ make the $t_{i}$ to go to either one of the three
values $-1,0,1$.

Computing $<S_{i}>$ in terms of $t_{i}$ and $K$ for each one of the 27 possible
limit values of the $t_{i}$ set, we find 7 solutions for the $m_{i}$ which are
respectively,
\[
m_{i}=0, \,  i = 1, 2, 3.
\]

and

\[
m_{i}=m_{j}=-m_{k} = \pm 1.
\]

To determine the actual minimum at $T=0$ we compute the associated values
for free energy
$F=-k_B{T}\log Z$. The first solution $m_{1}=m_{2}=m_{3}=0$ yields,
\[
F = -\Frac JK \log ( (6  + 2 \exp(4K))\exp(-K))
\mathop{\longrightarrow}_{K \rightarrow - \infty}  1
\]
and for $m_{1}=m_{2}=1, m_{3}=-1$ we get,
\[
F = -\Frac JK \log \left ( 2 \exp(-K) \cos(2 \delta K) (3 + \exp(4K))\right )
\mathop{\longrightarrow}_{K \rightarrow - \infty}
1 - 2 \delta ,
\]
making the solution  $m_{1}=m_{2}=1, m_{3}=-1$ the minimum. However 
from Eqs. (38)
and (39) the two free energies of the ordered/disordered phases are expected to
become equal only at some non zero temperature, a little bit above zero
temperature, that is quite close to a critical point. It is coherent 
to the known
result of a phase transition for the Triangular
Ising Antiferromagnet at $T=0$ in agreement with the previous improved mean
field theory by Netz and Berker [10].

\section{The Triangular Ising Ferromagnet }

Coming back to the TIF, we can go further and evaluate the value of 
the critical
temperature $K_{0}$. At this stage it is worth to notice, all above results are
independent of the value of $\delta$ which accounts for the coupling 
to the mean
field loops.

Since  $1 \leq \Frac{3\exp(4K) + 1}{\exp(4K)+3} \leq 3$, when $K>0$,
from Eq. (\ref{K}) we obtain,
\[
\Frac 1{6\delta} \leq K_0 \leq \Frac 1{2 \delta}.
\]
In addition, in the limit of large $\delta$, we get,
\[
K_0 =
\Frac 1{2 \delta}- \Frac{2}{\delta^2}+ \Frac 3{4\delta^3}
-\Frac{29}{24}\Frac
1{\delta^{4}}+{\cal O}\left (\Frac 1{\delta^5}\right )\ .
\]
To get a numerical estimate of the ferromagnetic
critical temperature $K_{0}$ requires to have the
  $\delta$ value.

 From Eq. (\ref{ham}) a straightforward arithmetic leads to
$\delta =\frac{q-2}{2}=2$ since $2$ nn are treated
exactly within the fluctuating loop out of the $6$ triangular n n.
Plugging then,
$\delta =2$ into Eq. (\ref{K}) yields $K_{0}= 0.1772$.
It is rather far from the exact numerical estimate $K_C^e=0.2746$ [14].
In comparison, a usual mean field gives $K_{0}= \frac{1}{6}=0.1667$, while
for Bethe it is $K_{0}= \tanh^{-1} (\frac{1}{5})=0.2027$.

\section{Conclusion}

In conclusion, we have showed that the very simple and generic mean field
Loopwise Scheme, proposed by Galam [7], is able to solve exactly the
Triangular Ising Antiferromagnet. Without any adjustable
parameter it recovers the exact Wannier argument of no ordering
at $T\neq 0$ and a transition at $T=0$ [10].
 From the same Equations the Triangular Ising
Ferromagnet is also solved simultaneously. A phase transition is obtained into
a ferromagnetic phase at a non-zero critical temperature

Moreover, contrary to the Bethe scheme, it preserves the
initial lattice symmetry, yet going beyond the one-site Weiss
approach. It
also  yields no transition for Ising hypercubes at $d=1$ with a lower critical
dimension of
$d_l=\frac{1+\sqrt 5}{2}$.

The Loopwise Scheme should allow a
new solving of a very large class of physical systems, in particular
random systems with frustration. For future work we consider to apply it first
to the Triangular Ising Antiferromagnet in a finite field and then on 
the stacked 3D
version of it. Application to the Random Field Ising model should also be done.


\subsection*{Acknowledgments.}
We would like to thank {\bf Y. Shapir} and {\bf R.  Netz} for 
stimulating discussion
on the manuscript.

\newpage
\subsection*{References}
\begin{enumerate}
\item
{\sf R. K. Pathria}, Statistical Mechanics, Pergamon Press (1972)
\item
{\sf Sh-k Ma},  {\em Modern Theory of Critical Phenomena}, The
Benjamin Inc.: Reading MA (1976)
\item
{\sf F. Y. Wu}, Rev. Mod. Phys. \underline {54},
235 (1982)
\item
{\sf P. Weiss}, J. Phys. Radium, Paris \underline {6}, 667 (1907)
\item
{\sf H. A. Bethe}, Proc. Roy. Soc. London A\underline {150}, 552 (1935)
\item
{\sf M. Suzuki},  Prog. Theor. Phys. \underline {42}  1086-1097 (1969)

\item
{\sf  S. Galam}, Phys. Rev. B\underline {54}, 15991 (1996)
\item
{\sf  S. Galam}, J. Appl. Phys. B\underline {87}, 7040 (2000)
\item
{\sf  G. Toulouse}, Commun. Phys.\underline {2}, 115 (1977)
\item
{\sf G. H. Wannier}, Phys. Rev. \underline {79}, 357 (1950)
\item
{\sf R. R. Netz and A. N. Berker}, Phys. Rev. Lett. \underline {6},
1377 (1991)
\item
{\sf J. R. Banavar, M. Cieplak, and A. Maritan}, Phys. Rev. Lett.
\underline {67}, 1807 (1991) and Reply by {\sf R. R. Netz and A. N. Berker}
\item
{\sf J. Monroe}, Physica A \underline {256}, pages 217 (1998)
\item
{\sf  J. Adler}, in ``Recent developments in computer simulation studies
in Condensed matter physics'', VIII, edited by D. P. Landau,
Springer (1995)

\end{enumerate}
\end{document}